\documentclass[journal]{IEEEtran}
\usepackage{graphicx}
\usepackage{color}
\usepackage{fancyhdr}
\usepackage[cmex10]{amsmath}
\usepackage{mdwmath}
\usepackage{amssymb}
\usepackage{textcomp}
\usepackage{times}
\usepackage{multicol}
\usepackage{float}
\usepackage[numbers,sort&compress]{natbib}

\usepackage{color}
\usepackage{algorithm} 
\usepackage{algorithmic} 
\usepackage{multirow}

\ifCLASSINFOpdf
\else
\fi
\begin{document}
%
\title{QoS-distinguished Achievable Rate Region for High Speed Railway Wireless Communications}
%
%


\author{Tao~Li,~Pingyi~Fan,~\IEEEmembership{Senior~Member,~IEEE}, Ke Xiong and K. B. Letaief, ~\IEEEmembership{Fellow,~IEEE}
\thanks{
T. Li and P. Fan are with Tsinghua National Laboratory for Information Science and Technology (TNList), and Department of Electrical Engineering, Tsinghua University, Beijing, China, 100084 (e-mail: \{litao12@mails, fpy@\}.tsinghua.edu.cn).

K. Xiong is with the School of Computer and Information Technology and the State Key Laboratory of Rail Traffic Control and Safety, Beijing Jiaotong University, Beijing, China, 100044 (e-mail:kxiong@bjtu.edu.cn).

K. B. Letaief is with Department of Electronic and Computer Engineering, Hong Kong University of Science and Technology, (e-mail: eekhaled@ece.ust.hk).
}
}

\maketitle
\thispagestyle{empty}

\begin{abstract}

In high speed railways (HSRs) communication system, when a train travels along the railway with high velocity, the wireless channel between the train and base station varies strenuously, which makes it essential to implement appropriate power allocations to guarantee system performance. What's more, how to evaluate the performance limits in this new scenario is also needed to consider.

To this end, this paper investigates the performance limits of wireless communication in HSRs scenario. Since the hybrid information transmitted between train and base station usually has diverse quality of service (QoS) requirements, QoS-based achievable rate region is utilized to characterize the transmission performance in this paper. It is proved that traditional ergodic capacity and outage capacity with unique QoS requirement can be regarded as two extreme cases of the achievable rate region proposed in this paper. The corresponding optimal power allocation strategy is also given to achieve the maximal boundary of achievable rate region.

Compared with conventional strategies, the advantages of the proposed strategy are validated in terms of green communication, namely minimizing average transmit power.
Besides, the hybrid information transmission in a non-uniform generalized motion scenario is analyzed to confirm the robust performance of proposed strategy. The performance loss caused by non-uniform motion compared with that in uniform motion is also indicated, where a deterministic worst case for instantaneous speed realization is proposed to serve as the lower bound for system performance.
\end{abstract}


%
\IEEEpeerreviewmaketitle

\section{Introduction}
%
%
%
%

\IEEEPARstart{H}{igh} speed railways (HSRs) have been developed rapidly in recent years, especially in China, where more than ten thousand kilometers of HSRs have been built by the end of $2013$. Consequently, high mobility wireless communication attracts much more attention than ever before \cite{Zhou_1}. Many key technologies of wireless communication need to be reconsidered in high mobility scenarios, such as channel estimation, synchronization, multiple antenna technique, resource allocation, etc \cite{Dai_2,Yang_3,Wang_4,Dong_5}.
For instance, in \cite{Dai_2}, it proposed a robust time-frequency training algorithm for OFDM signal block in high mobility environment. A novel Doppler frequency offset estimation method was proposed by \cite{Yang_3} for DVB-T system in HSRs. The work in \cite{Wang_4} investigated the effect of distributed antenna techniques on the hand-off frequency of system. The upper and lower bounds of channel capacity of high mobility wireless channel have been analyzed in \cite{Dong_5}.

Among them, power allocation, one of the most important methods to guarantee reliability and efficiency of information transmission, is the main problem that this paper will concentrate on.
Of course, fixed transmit power is the simplest strategy for power allocation in HSRs. In \cite{Zhang_6}, it analyzed the system service in HSRs under fixed transmit power, where some base station deployment strategies were obtained. However, it has been already demonstrated that fixed transmit power is not a good choice unless in time-invariant additive white Gaussian noise channel (AWGN) \cite{Gallager_7,Biglieri_20}.
From a prospective of maximizing throughput, namely the ergodic capacity, water-filling algorithm is optimal in fading channel \cite{Tse_8}. On the other hand, when transmission delay is also taken into account, the concept of outage capacity is proposed for measuring transmission performance, and the corresponding optimal power allocation strategy is channel inversion algorithm \cite{Goldsmith_9}. Effective capacity is another powerful method to describe the channel limits for delay limited information transmission \cite{Wu_10,Efazati_11}. In \cite{Luo_12}, it investigated the maximal ergodic throughput when outage probability is constrained. Besides, fairness among multiple users is also an important factor that needs to take into account \cite{Dong_13,Shi_14}, where tradeoff relationship between efficiency and fairness can serve as an important performance metric.

Considering the specific characteristic of HSRs, under the widely used relay-aided two-hop system structure proposed in \cite{Zhou_1}, the information transmitted between base station (BS) and access point (AP) is a mixed version that come from many different passengers on the train, namely a group of information flows with diverse quality of service (QoS) requirements, such as different data rates and maximal tolerant delays. For simplifying discussion, we call this kind of information as hybrid information in the sequel. According to the literature \cite{Tse_8,Hanly_15}, information flows can be at least divided into two categories based on delay requirement: delay-sensitive stream and delay-insensitive stream. It is obvious that traditional power allocation strategies optimized with respect to simple parameter are not appropriate for hybrid information case in HSRs scenario. What's more, we need to develop new metric to characterize the performance limits of wireless channel in this condition instead of traditional metrics, such as ergodic capacity and outage capacity.

According to above considerations, this paper focuses on the information transmission between BS and AP, and attempts to explore the performance limits of wireless channel in HSRs scenarios. With the assumption that high speed train travels along a line railway with a constant velocity, this paper employs a point-to-point model under which delay-sensitive and delay-insensitive information streams are transmitted simultaneously. In order to investigate the tradeoff between the rates of these two concurrent streams, QoS-distinguished achievable rate region is utilized to characterize the hybrid information transmission performance limits of wireless communication system in HSRs. It is found that conventional ergodic capacity and outage capacity can be regarded as two extreme cases of the proposed achievable rate region. The corresponding optimal power allocation strategy, which can match with hybrid information transmission very well, is derived to achieve the maximal boundary of achievable rate region. In order to validate the results proposed by this paper, we compare the new strategy with other traditional ones from a perspective of green communication, namely minimizing average transmit power. Among which, an interesting observation is derived that the average transmit power is independent with train's velocity on the condition that QoS requirements have been given, which is very helpful for the feasibility of applying this algorithm in HSRs. Lastly, the transmission performance in a non-uniform generalized motion scenario is discussed to consider the robust performance of the proposed algorithm. A worst case for instantaneous speed realization is provided as the lower bound for system performance. And the performance loss resulted from non-uniform motion is indicated through simulation results compared with that in uniform motion scenario. Besides, it is worth noting that the delay mainly considered in this paper is caused by time variant characteristic of wireless channel. The coding delay which was discussed in \cite{Berry_16} is beyond the scope of this paper.

The rest of this paper is organized as follows. Two-hop system structure and channel model for wireless communication system in HSR scenarios are introduced in Section II. Diverse QoS requirements of multiple information flows, especially delay requirement, are presented in Section III. The QoS-distinguished achievable rate region is discussed in Section IV and some simulation results are also given to validate its optimality. From a perspective of green communication, QoS-distinguished power allocation algorithm is analyzed in Section V. Then, the robustness of the proposed strategy in a non-uniform generalized motion scenario is discussed in Section VI. At last, some conclusions are given in Section VII.

\section{Preliminary}

\subsection{System Structure}

\begin{figure}[!t]
\centering
\includegraphics[width=3.3 in]{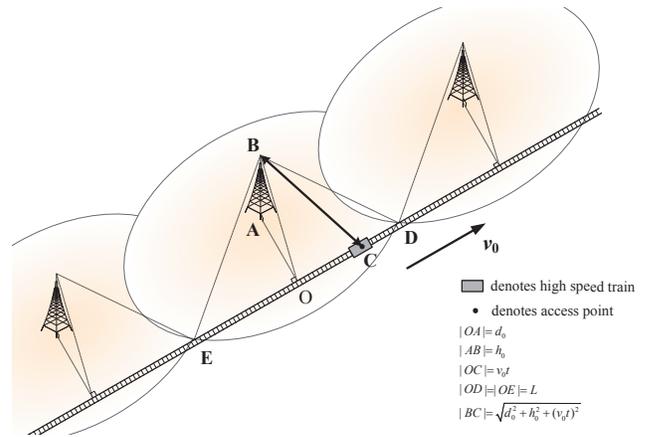}
\caption{The structure diagram of wireless communication system in high speed railways between high speed train and base station.}
\label{fig:Base_Station_Deployment}
\end{figure}

A typical system structure diagram for high speed railway wireless communication system is shown in Fig. \ref{fig:Base_Station_Deployment}, where BSs are uniformly deployed along the railway with equal interval. The distance between BS and railway is $d_0$, the height of antenna equipped at each BS is $h_0$, and the coverage radius of each BS along railway is $L$. In Fig. \ref{fig:Base_Station_Deployment}, it is assumed that a high speed train is traveling along a line railway with a constant velocity, which is denoted as $v_0$. The performance in non-uniform motion scenario will be explored in detail in Section VI.

In order to avoid serious penetration loss caused by metal carriage and implement group handoff between adjacent cells, users on the train usually communicate with base station (BS) under the help of access point (AP), which is usually installed on the roof of train. Since information transmission process between AP and users on the train is the same as traditional network, some traditional technology, such as WIFI, can solve this problem very well. Thus, this paper focuses on the information transmission between the mobile AP and BS. Provided that AP is only equipped with one antenna, so the train can be regarded as a mobile mass point on the railway.
Besides, it is assumed that the system time $t$ is equal to $0$ when the train passes the point $O$ as shown in Fig. \ref{fig:Base_Station_Deployment}. Since transmission process along time is periodic based on the above assumptions, without loss of generality, we concentrate on discussing the information transmission problem under the coverage of one BS, which can reflect all the characteristics of this system. That is to say, the system time range that will be taken into account is $t\in [-\frac{L}{v_0},\frac{L}{v_0}]$ without any other declaration in the sequel. Under above assumptions, the real-time location of train is $v_0t$ when system time is $t$ $(t\in [-\frac{L}{v_0},\frac{L}{v_0}])$, and the corresponding transmission distance between BS antenna and AP antenna can be expressed as $d(t)=\sqrt{d_0^2+h_0^2+(v_0 t)^2}$.

\subsection{Channel Model}

Since AP has multiplexed the information flows from different users on the train, the transmission process between AP and BS is equivalent to a point-to-point transmission process. Let $x(t)$ and $y(t)$ be input signal and output signal, respectively. Generally, $x(t)$ can be assumed as zero mean and unit variance. The available frequency bandwidth of system is denoted as $B$. Then, the baseband-equivalent instantaneous-time model for the channel between AP and BS can be expressed as following
\begin{equation}\label{equ:channel model}
  y(t)=\sqrt{\frac{G h(t) p(t)}{d(t)^\alpha}} x(t)+n(t).
\end{equation}
where $G$ denotes antenna gain coefficient, $\alpha$ denotes path loss exponent, $h(t)$ denotes channel fading coefficient, and $p(t)$ denotes transmit power at system time $t$. $n(t)$ is additive complex cycle symmetric Gaussian noise with zero mean and variance $\sigma_0^2$.

Thus, the corresponding instantaneous information transmission capacity at system time $t$ is expressed as follows, which can be achieved by Gaussian input signal.
\begin{equation}\label{equ:instantaneous information capacity}
  R(t)=B \log_2\Big(1+\frac{G h(t)p(t)}{d(t)^\alpha \sigma_0^2}\Big).
\end{equation}

It is assumed that $h(t)$ is a constant value in the following discussion, such as $h(t)=1$. Two reasons motivate us to employ this channel model: 1) line-of-sight (LOS) component is always dominant in HSRs scenario since majority of HSRs is running over a viaduct in plain area where is in the absence of scatterer \cite{Liu_17,Li_18,Luo_19}. 2) the main problem this paper concentrates on is power allocation for large-scale fading, which changes not so fast as small-scale fading that power allocation module can track it.

\section{QoS Requirements in HSRs Scenario}

Usually, information flows from different users (or applications) have different QoS requirements, such as data rate, maximal tolerant delay and delay violation probability. With the help of enough buffer at transmitter, delay violation only occurs when transmission delay exceeds the maximal tolerant value. As we know, regardless of the delay requirement, water-filling algorithm is optimal for power allocation in terms of maximizing data rate, where ergodic capacity is an effective metric for measuring system performance. On the other hand, if the service is sensitive to delay, widely used channel inversion algorithm is optimal, where outage capacity is more approach for system performance. They are equivalent when and only when channel environment is time-invariant, such as AWGN channel. As a consequence, the optimality of power allocation algorithm depends on the specific QoS requirements. Consider that the information flow transmitted between AP and BS is a hybrid version that contains both delay-insensitive and delay-sensitive services, traditional unique ergodic capacity and outage capacity are not sufficient to describe the channel transmission performance in this new scenario.

Assuming $N$ different information flows needed to be transmitted between the mobile AP and BS as shown in Fig. \ref{fig:Base_Station_Deployment}, we denote the vectors of data rate requirements and maximal tolerant delays as $\mathbf{R}=(R_1,R_2...R_N)$ and $\mathbf{\tau}=(\tau_1,\tau_2...\tau_N)$, respectively. For the $i$-th information flow, if $\tau_i \geq \frac{2L}{v_0}$ (which is the operation period of HSR), it can be viewed as delay-insensitive information since one operation period is enough to complete the transmission without delay violation. Otherwise, if $\tau_i<\frac{2L}{v_0}$, we refer it to as delay-sensitive information, which needs real-time transmission in order to avoid delay violation. Thus, the hybrid information flows between AP and BS can be divided into two categories based on transmission delay requirements, i.e., delay-insensitive stream and delay-sensitive stream.
Let the sums of data rate requirements for delay-insensitive and delay-sensitive information streams among the $N$ information flows be $R_{di}$ and $R_{ds}$, respectively. Under above assumptions, they can be given by
\begin{equation}\label{equ:two information rates}
  R_{di}=\sum_{\{i|\tau_i \geq \frac{2L}{v_0}\}} R_i,\,\,\,\,\,\,\,\,\,\, R_{ds}=\sum_{\{i|\tau_i < \frac{2L}{v_0}\}} R_i
\end{equation}

For hybrid information transmission with diverse QoS requirements, the data rate pair $(R_{di},R_{ds})$ can used as a new metric, taking placing of traditional capacity $C$, to provide more insight of channel characteristic. When the average transmit power is constrained at transmitter, it is apparent that there exists a tradeoff between $R_{di}$ and $R_{ds}$. Thus, we define \emph{QoS-distinguished achievable rate region} $\{(R_{di},R_{ds})\}$ to characterize the system performance limits of wireless communication channel in HSR scenario.

\emph{\textbf{Definition 1: QoS-distinguished achievable rate region $\{(R_{di},R_{ds})\}$} is defined as the set composed of all feasible delay-insensitive information rate and delay-sensitive information rate pair $(R_{di},R_{ds})$ under the average transmit power constraint.}

\section{Performance Limits Analysis in HSRs}

It has indicated that achievable rate region $\{(R_{di},R_{ds})\}$ is more fit for measuring transmission performance of hybrid information with diverse QoS requirements in HSRs scenario. This section will investigate the information transmission limits of the system shown as in Fig. \ref{fig:Base_Station_Deployment}. Our goal is to obtain an explicit closed-form expression of achievable rate region $\{(R_{di},R_{ds})\}$. To do so, two extreme cases in which either $R_{di}$ or $R_{ds}$ is equal to zero will be considered firstly, to serve as two baselines for hybrid information transmission. Then, the expression of QoS-distinguished achievable rate region in HSRs will be derived with the help of conditional capacity function. Moreover, the corresponding optimal power allocation strategy which can achieve the maximal boundary of $\{(R_{di},R_{ds})\}$ region will also be given.

\subsection{Delay-sensitive Information Transmission}

Firstly, we consider the extreme case that $R_{di}$ is equal to zero, it means all information to transmit is delay-sensitive. The problem degrades to traditional zero-outage capacity problem, which corresponds to the maximal achievable value of $R_{ds}$, denoted as $R_{ds}^{max}$. Provided that average transmit power is $P_0$ and the train travels along a line railway with constant velocity $v_0$, $R_{ds}^{max}$ in this case should be modeled as the following optimization problem.
\begin{subequations}\label{equ:delay-sensitive case capacity}
\begin{align}
  & R_{ds}^{max}=\max \limits_{p(t)} \{ \min \{R(t)|t\in \big[-\frac{L}{v_0},\frac{L}{v_0}\big] \} \}  \tag{4} \\
  & s.t. \,\,\,\,\,\,\,\,\, \frac{v_0}{2L}\int_{-\frac{L}{v_0}}^{\frac{L}{v_0}}p(t)dt \leq P_0 \label{equ:delay-sensitive case capacity a}
\end{align}
\end{subequations}
where $R(t)$ is the instantaneous transmit rate at system time $t$ shown in (\ref{equ:instantaneous information capacity}). The constraint (\ref{equ:delay-sensitive case capacity a}) denotes the limited average transmit power constraint.

By solving the problem in (\ref{equ:delay-sensitive case capacity}), the expression of $R_{ds}^{max}$ and the corresponding power allocation strategy can be concluded as Lemma 1. 

\emph{\textbf{Lemma 1:} The maximal achievable value of $R_{ds}$ is}
\begin{equation}\label{eqn:maximal achievable value of R-ds}
  R_{ds}^{max}=B \log_2\Big(1+\frac{G P_0 \cdot \frac{2L}{v_0}}{\sigma_0^2 \int_{-\frac{L}{v_0}}^{\frac{L}{v_0}}d(t)^\alpha dt}\Big)
\end{equation}
\emph{where optimal power allocation strategy $p_{ci}^\ast(t)$ is}
\begin{equation}\label{equ:delay-sensitive case optimal power allocation}
  p_{ci}^\ast(t)=\frac{d(t)^\alpha \sigma_0^2}{G}\big(2^{\frac{R_{ds,max}}{B}}-1\big), t\in \big[-\frac{L}{v_0},\frac{L}{v_0}\big]
\end{equation}
\begin{proof}
Substituting (\ref{equ:instantaneous information capacity}) into (\ref{equ:delay-sensitive case capacity}), based on the optimization method in \cite{Goldsmith_9}, the expression of $p^\ast(t)$ can be obtained, which is shown in (\ref{equ:delay-sensitive case optimal power allocation}). Then, substituting (\ref{equ:delay-sensitive case optimal power allocation}) into (\ref{equ:delay-sensitive case capacity a}), the result in (\ref{eqn:maximal achievable value of R-ds}) can be obtained through some manipulations.
\end{proof}

\subsection{Delay-insensitive Information Transmission}

When $R_{ds}$ is $0$, it means that all the transmitted information is insensitive to transmission delay. The maximal achievable value of $R_{di}$, denoted as $R_{di}^{max}$, is equivalent to ergodic capacity. Therefore, $R_{di}^{max}$ can be modeled as the solution to the following problem.
\begin{subequations}\label{equ:delay-insensitive case capacity}
\begin{align}
  & R_{di}^{max}=\max \limits_{p(t)} \Big\{ \frac{v_0}{2L}\int_{-\frac{L}{v_0}}^{\frac{L}{v_0}}R(t) dt\Big\}  \tag{7}\\
  & s.t. \,\,\,\,\,\, \frac{v_0}{2L}\int_{-\frac{L}{v_0}}^{\frac{L}{v_0}}p(t)dt \leq P_0 \label{equ:delay-insensitive case capacity a}
\end{align}
\end{subequations}

\emph{\textbf{Lemma 2:} When $R_{ds}$ is $0$, the maximal achievable value of $R_{di}$ is}
\begin{equation}\label{eqn:maximal achievable value of R-di}
  R_{di}^{max}=\frac{v_0}{2L}\int_{-\frac{L}{v_0}}^{\frac{L}{v_0}} B \log_2\Big(1+\frac{G p_{wf}^\ast(t)}{d(t)^\alpha \sigma_0^2}\Big) dt
\end{equation}
\emph{where the optimal allocation strategy is}
\begin{equation}\label{equ:delay-insensitive case optimal power allocation}
  p_{wf}^\ast(t)=\frac{B}{\lambda_0}-\frac{d(t)^\alpha \sigma_0^2}{G}, t\in \big[-\frac{L}{v_0},\frac{L}{v_0}\big].
\end{equation}
\emph{where $\lambda_0$ is a constant value determined by average transmit power constraint in (\ref{equ:delay-insensitive case capacity a}).}

\subsection{Hybrid Information with Diverse QoS Requirements}

For the general case in which neither $R_{di}$ nor $R_{ds}$ is zero, it is obvious that the rate pair $(R_{di},R_{ds})$ is located within the two-dimensional region $[0,R_{di}^{max}]\times[0,R_{ds}^{max}]$. Two objective functions $R_{di}$ and $R_{ds}$ should be considered together, which is a typical multi-objective optimization problem. That is to say, the main task in this subsection is to maximize the region $\{(R_{di},R_{ds})\}$ under the total average transmit power constraint.

Since the QoS requirement of delay-sensitive information stream is more rigorous than that of delay-insensitive information stream, $R_{ds}$ should be given priority in time-variant scenario.
Thus, in order to obtain the boundary line of achievable rate region $\{(R_{di},R_{ds})\}$, the basic idea of our method to explore the $\{(R_{di},R_{ds})\}$ region is to maximize the delay-insensitive average information rate $R_{di}$ with the limited average transmit power constraint after the delay-sensitive average information rate $R_{ds}$ has been satisfied. In order to make the discussion in the sequel more clear, we give the definition of conditional capacity function as follows.

\emph{\textbf{Definition 2: conditional capacity function $C_{R_{ds}}$} is defined as the maximal achievable value of delay-insensitive average information rate $R_{di}$ when transmit power is constrained and delay-sensitive information rate is $R_{ds}$.}

According to the Definition 2, the QoS-distinguished achievable rate region $\{(R_{di},R_{ds})\}$ can be expressed as
\begin{equation}\label{equ:solution for information rate region}
  \{(R_{di},R_{ds})|0\leq R_{ds} \leq R_{ds}^{max},0\leq R_{di} \leq C_{R_{ds}} \}
\end{equation}

Assuming that the average transmit power is $P_0$, when the delay-sensitive transmission rate requirement is $R_{ds}$, the optimization problem for $C_{R_{ds}}$ can be model as
\begin{subequations}\label{equ:optimization problem for power allocation}
\begin{align}
   C_{R_{ds}}&= \max \limits_{p(t)} \Big\{ \frac{v_0}{2L}\int_{-\frac{L}{v_0}}^{\frac{L}{v_0}}[R(t)-R_{ds}]dt\Big\}   \tag{11}\\
   s.t.\,\,\,\,\,& \frac{v_0}{2L}\int_{-\frac{L}{v_0}}^{\frac{L}{v_0}}p(t)dt \leq P_0 \label{equ:optimization problem for power allocation a} \\
   & R(t)=B \log_2\big(1+\tfrac{G p(t)}{d(t)^\alpha \sigma_0^2}\big)  \label{equ:optimization problem for power allocation b} \\
   & R(t)\geq R_{ds}\geq 0,t\in [-\tfrac{L}{v_0},\tfrac{L}{v_0}]  \label{equ:optimization problem for power allocation c}
\end{align}
\end{subequations}
where the constraint (\ref{equ:optimization problem for power allocation a}) denotes average transmit power constraint, the constraint (\ref{equ:optimization problem for power allocation b}) denotes instantaneous information transmission throughput at system time $t$, and the inequality in (\ref{equ:optimization problem for power allocation c}) denotes the constraint of real-time information data rate requirement.

\emph{\textbf{Proposition 1:} When transmit power is $P_0$ and data rate requirement for delay-sensitive information is $R_{ds}$, the conditional capacity function $C_{R_{ds}}$ can be expressed as}
\begin{equation}\label{equ:maximum non-real-time rate}
  C_{R_{ds}}=\frac{v_0}{2L}\int_{-\frac{L}{v_0}}^{\frac{L}{v_0}}\Big[B \log_2\big(1+\frac{G p_{hb}^\ast(t)}{d(t)^\alpha \sigma_0^2}\big)-R_{ds}\Big]dt
\end{equation}
\emph{where the optimal power allocation algorithm $p_{hb}^\ast(t)$ is}
\begin{equation}\label{equ:optimal power allocation solution}
  p_{hb}^\ast(t)=\Big(\frac{d(t)^\alpha \sigma_0^2}{G}(2^{\frac{R_{ds}}{B}}-1),\frac{B}{\lambda_0'}-\frac{d(t)^\alpha \sigma_0^2}{G}\Big)^+, t\in \big[-\frac{L}{v_0},\frac{L}{v_0}\big]
\end{equation}
\emph{where operation $(x_1,x_2)^+$ denotes the larger one between $x_1$ and $x_2$, $\lambda_0'$ is a constant value which is determined by the average power constraint in (\ref{equ:optimization problem for power allocation a}).}
\begin{proof}
See the Appendix A.
\end{proof}

For description convenience, the optimal power allocation strategy described in (\ref{equ:optimal power allocation solution}) is called as hybrid allocation algorithm. Delay-sensitive information transmission requirement is considered separately with the delay-insensitive information in hybrid allocation algorithm in (\ref{equ:optimal power allocation solution}), which makes system more efficient in the respect of hybrid information transmission. Besides, according to the (\ref{equ:optimal power allocation solution}), it can be observed that $p_{hb}^\ast(t)$ will degrade to channel inversion algorithm $p_{ci}^\ast(t)$ when $R_{ds}$ is equal to $R_{ds,max}$ and $p_{hb}^\ast(t)$ will degrade to water-filling algorithm $p_{wf}^\ast(t)$ when $R_{ds}$ is equal to $0$. Consequently, channel inversion algorithm and water-filling algorithm can be regarded as two extreme cases of the results in (\ref{equ:optimal power allocation solution}). Thus, the QoS-distinguished achievable rate region $\{(R_{di},R_{ds})\}$, bridging traditional outage capacity and ergodic capacity, is a more efficient metric for transmission performance of hybrid information flows with diverse QoS requirements.

\subsection{Simulation Results and Comments}

\begin{figure}[!t]
\centering
\includegraphics[width=3.3 in]{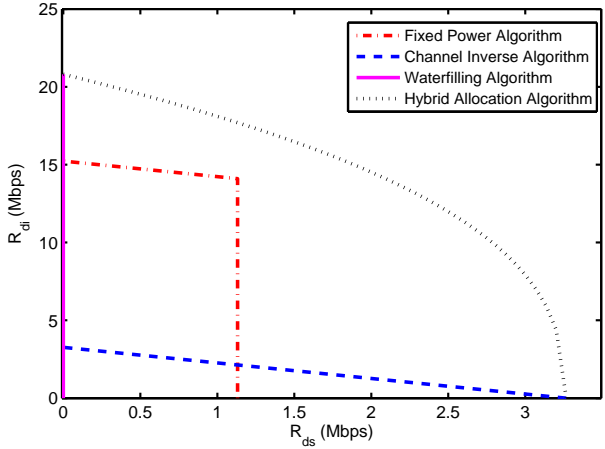}
\caption{The achievable rate regions under four different power allocation strategies when average transmit power $P_0$ is $30$ dBm, which contains: fixed power algorithm, channel inversion algorithm, water-filling algorithm and hybrid allocation algorithm.}
\label{fig:Simulation_for_Two_Rate_Region}
\end{figure}

Some simulation results will be given in this subsection to illustrate the superiority of the hybrid allocation algorithm proposed by this paper compared with other conventional algorithms.
For the system shown in Fig. \ref{fig:Base_Station_Deployment}, it is assumed that the distance between BS and railway $d_0$ is $2m$, the height of antenna at BS $h_0$ is $10m$ and the coverage radius of each BS along railway $L$ is $500m$. The velocity of high speed train $v_0$ is $100m/s$ (e.g. $360km/h$). The antenna gain $G$ is $1$, additive channel noise power $\sigma_0^2$ is $0$ dBm, path loss exponent $\alpha$ is $2$ and the bandwidth of frequency $B$ is $20$ MHz.
We will compare the transmission performance of different power allocation strategies in terms of achievable rate region with average transmit power constraint. Fig. \ref{fig:Simulation_for_Two_Rate_Region} plots the simulation results when the average transmit power $P_0$ is $30$ dBm, which presents the performance of four different algorithms: fixed power algorithm, channel inversion algorithm, water-filling algorithm and hybrid allocation algorithm. It can be seen from Fig. \ref{fig:Simulation_for_Two_Rate_Region} that hybrid allocation algorithm is superior to any other algorithms. Fixed power algorithm is not optimal for all cases due to the time-varying characteristic of channel, which accords with the intuition. Channel inversion algorithm is optimal if and only if $R_{di}=0$, and water-filling algorithm is optimal if and only if $R_{ds}=0$. Hybrid allocation algorithm bridges the gap between channel inversion algorithm and water-filling algorithm, and provides the largest achievable region $\{(R_{di},R_{ds})\}$. The $\{(R_{di},R_{ds})\}$ in other traditional algorithms is just a subset of that in hybrid allocation algorithm.

It has been demonstrated that hybrid algorithm for power allocation is the most powerful strategy when both delay-sensitive and delay-insensitive information streams are simultaneously transmitted over the same wireless channel. Then, as a comparison, we consider another case in which two sub-channels are available, denoted as \emph{sub-channel A} and \emph{sub-channel B}, respectively. Each of the channels has the same parameters as stated for Fig. \ref{fig:Simulation_for_Two_Rate_Region}. In fact, we can also construct two sub-channels by dividing one channel into two through multiplex technique in certain domain, such as frequency division multiplex (FDM). In this case, a simple but widely used scheduling is to transmit different information streams over different sub-channels, respectively, which is called as separate transmission schedule. For example, \emph{sub-channel A} loads the delay-sensitive information stream while \emph{sub-channel B} loads the delay-insensitive information stream. That is, each channel is loaded with service with unique QoS requirement, and traditional optimal power allocation strategy is available once again. Channel inversion algorithm is adopted for power allocation in \emph{sub-channel A} while water-filling algorithm is used for power allocation \emph{sub-channel B}. According to above considerations, a basic problem will be proposed naturally: compared with the separate transmission schedule as introduced above, whether simultaneous transmission schedule under hybrid allocation algorithm can obtain a better performance?

\begin{figure}[!t]
\centering
\includegraphics[width=3.3 in]{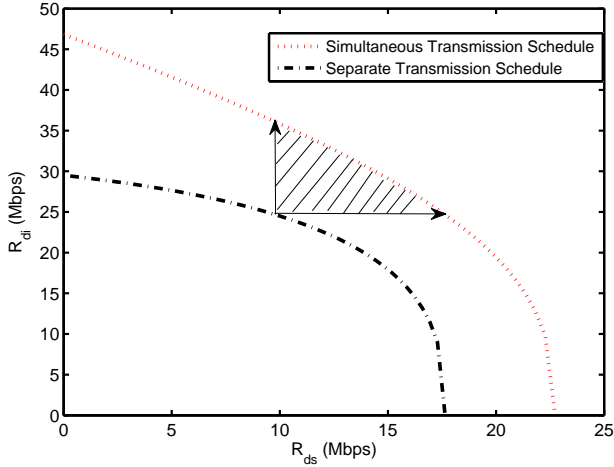}
\caption{The achievable rate regions under simultaneous transmission schedule and separated transmission schedule, respectively, when there are two available channels and the total average transmit power $P_0$ is $40$ dBm.}
\label{fig:Simulation_for_Two_Strategies}
\end{figure}

Fig. \ref{fig:Simulation_for_Two_Strategies} plots the simulation results under simultaneous transmission schedule and separated transmission schedule in terms of achievable rate region when there are two available sub-channels and the total average transmit power $P_0$ is $40$ dBm. In simultaneous transmission schedule, power allocation is optimized based on QoS requirements by the hybrid allocation algorithm proposed by this paper. On the other hand, in separate transmission schedule, water-filling algorithm is used for the channel that loads delay-insensitive information flow while channel inversion algorithm is used for the channel that loads delay-sensitive information flow.
It can be seen from Fig. \ref{fig:Simulation_for_Two_Strategies} that the performance of simultaneous transmission schedule is apparently better than that of separate transmission schedule in terms of achievable rate region. For example, $10$ Mbps delay-sensitive information and $25$ Mbps delay-insensitive information can be transmitted under separate transmission schedule between BS and AP at the same time. Whereas, if simultaneous transmission schedule is employed, the system can work under arbitrary possible point inside the shaded area shown in Fig. \ref{fig:Simulation_for_Two_Strategies}, in which both $R_{di}$ and $R_{ds}$ are larger than these in separate transmission schedule, to achieve a better transmission performance. Specifically, $7.5$ Mbps more delay sensitive information or $12$ Mbps delay-insensitive information can be transmitted with the same average transmit power constraint. Even for the extreme case, such as $R_{ds}=0$, simultaneous transmission schedule is better than separate schedule, the performance gain of which is nearly $57\%$ with respect to the throughput in separate schedule. It is because that all degree of freedom in wireless channel can be made full use in simultaneous transmission schedule, which is great helpful to improve system performance. In summary, simultaneous transmission schedule is a good choice even there are multiple available channels.

\section{QoS-distinguished Power Allocation for Green Communication}

\subsection{Minimizing Energy Consumption Algorithm}

For many practical applications, the data rate requirements of delay-insensitive information $R_{di}$ and delay-sensitive information $R_{ds}$ are usually prior known by transmitter and receiver. In this case, it is more meaningful to minimize the whole energy consumption of the whole system by distinguishing the different QoS requirements and matching the power allocation strategy with the characteristic of information flows. Let the required average transmit power be $P_{0}^{min}$, the optimization object is to minimize the average transmit power that is needed to support the data rate pair $(R_{di},R_{ds})$. Similar to the problem in (\ref{equ:optimization problem for power allocation}), it can be modeled as the following optimization problem on the condition that $R_{di}$ and $R_{ds}$ are given.
\begin{subequations}\label{equ:optimization problem for minimizing transmit power}
\begin{align}
   P_{0}^{min}& =\min \limits_{p(t)} \Big\{ \frac{v_0}{2L}\int_{-\frac{L}{v_0}}^{\frac{L}{v_0}}p(t)dt \Big\}   \tag{14}\\
   s.t.\,\,\,\,\,& R(t)=B \log_2\Big(1+\frac{G p(t)}{d(t)^\alpha \sigma_0^2}\Big), \, t\in \big[-\frac{L}{v_0},\frac{L}{v_0}\big] \label{equ:optimization problem for minimizing transmit power a} \\
   & 0 \leq R_{di} \leq \frac{v_0}{2L} \int_{-\frac{L}{v_0}}^{\frac{L}{v_0}}[R(t)-R_{ds}]dt \label{equ:optimization problem for minimizing transmit power b} \\
   & 0 \leq R_{ds} \leq R(t), \, t\in \big[-\frac{L}{v_0},\frac{L}{v_0}\big]  \label{equ:optimization problem for minimizing transmit power c}
\end{align}
\end{subequations}
where the constraint (\ref{equ:optimization problem for minimizing transmit power a}) denotes instantaneous transmit capacity limit. The constraints (\ref{equ:optimization problem for minimizing transmit power b}) and (\ref{equ:optimization problem for minimizing transmit power c}) denote delay-insensitive and delay-sensitive information data rate requirements, respectively.

\emph{\textbf{Proposition 2:} When information data rate requirements $(R_{di},R_{ds})$ are given, the optimal allocation strategy in terms of minimizing energy consumption is}
\begin{equation}\label{equ:optimal power allocation solution in terms of minimizing transmit power}
  p_{m}^\ast(t)=\Big(\frac{d(t)^\alpha \sigma_0^2}{G}(2^{\frac{R_{ds}}{B}}-1),\frac{B}{\lambda_0''}-\frac{d(t)^\alpha \sigma_0^2}{G }\Big)^+,t\in \big[-\frac{L}{v_0},\frac{L}{v_0}\big]
\end{equation}
\emph{where the constant value $\lambda_0''$ is determined by the following constraint}
\begin{equation}\label{eqn:delay insensitive rate constraint}
  R_{di}=\frac{v_0}{2L}\int_{-\frac{L}{v_0}}^{\frac{L}{v_0}}\Big[B \log_2\big(1+\frac{G p_{m}^\ast(t)}{d(t)^\alpha \sigma_0^2}\big)-R_{ds}\Big]dt
\end{equation}

As a result, the minimum transmit power $P_{0}^{min}$ can be expressed as:
\begin{equation}\label{eqn:minimum transmit power in Proposition 2}
  P_{0}^{min}=\frac{v_0}{2L}\int_{-\frac{L}{v_0}}^{\frac{L}{v_0}}p_m^\ast(t)dt
\end{equation}

The proof of Proposition 2 is similar to that of Proposition 1, so it is ignored here due to limit of space. Besides, based on the results above, we can give a specific algorithm for power allocation in terms of minimizing the average transmit power shown as Algorithm 1.
\begin{algorithm}[htb]
\caption{Calculating the minimal average transmit power that can support the QoS requirements.}
\label{alg:Framwork}
\begin{algorithmic}[1] 
\REQUIRE ~~\\ 
The data rate requirements pair $(R_{ds},R_{di})$;\\
The parameters for system deployment $d_0, h_0$ and $L$;
\ENSURE ~~\\ 
The minimal requirement average transmit power $P_{0}^{min}$;
\STATE set an original value for $\lambda_0''$;
\STATE calculate current data rate of delay-insensitive information $R_{di}(\lambda_0'')$ on condition of $\lambda_0''$ by Eqn. (\ref{equ:optimal power allocation solution in terms of minimizing transmit power}-\ref{eqn:delay insensitive rate constraint}); \label{algorithem_begin}
\IF{$|R_{di}(\lambda_0'')-R_{di}|<0.001$}
\STATE goto to Step (\ref{algorithm_end});
\ELSE
\IF{$R_{di}(\lambda_0'')>R_{di}$}
\STATE $\lambda_0''=\lambda_0''/(1+0.1)$;
\ELSE
\STATE $\lambda_0''=\lambda_0''*(1+0.07)$;
\ENDIF
\STATE return to Step (\ref{algorithem_begin});
\ENDIF
\STATE calculating minimal average transmit power $P_{0}^{min}$ by substituting the value of $\lambda_0''$ into Eqn. (\ref{eqn:minimum transmit power in Proposition 2}). \label{algorithm_end}
\end{algorithmic}
\end{algorithm}

\subsection{Performance Analysis}

Assuming the system parameters are the same as those in Fig.\ref{fig:Simulation_for_Two_Rate_Region}, Table I gives the minimum required average transmit power under four different power allocation algorithms when data rate pair $(R_{di},R_{ds})$ are shown in the first column. In the table, the abbreviations of FPA, WFA, CIA and HAA denote fixed power algorithm, water-filling algorithm, channel inversion algorithm and hybrid allocation algorithm (QoS-based allocation algorithm), respectively. The minimum required average transmit power that corresponds to the optimal allocation algorithm in each considered instance $(R_{di},R_{ds})$ are highlighted with red colour.

It can be observed that HAA is the best strategy in all considered instance $(R_{di},R_{ds})$ in terms of minimizing transmit power, namely, obtaining the highest efficiency of energy. Since the delay-insensitive information is also regarded as delay-sensitive information under CIA, the required transmit power will stay the same on the condition that the sum of $R_{di}$ and $R_{ds}$ is fixed, which is shown in the forth column of Table I. Compared with HAA, the performance of WFA deteriorates very heavily unless $R_{ds}=0$ while the performance of CIA is also very bad unless $R_{di}=0$. It is because that both WFA and CIA can only satisfy unique QoS requirement at a time, they will become mismatch with respect to hybrid information transmission. Besides, FPA is not optimal in all the cases that has been considered in Table I, which agrees with the results obtained in the previous section.

After the optimal power allocation algorithm for hybrid information transmission in HSRs in terms of minimizing average transmit power has been given in (\ref{equ:optimal power allocation solution in terms of minimizing transmit power}), let's discuss the robust performance of the proposed algorithm, such as the relationship between average transmit power and train's velocity. One interesting observation can be drawn from the results, which is expressed as Proposition 3.

\emph{\textbf{Proposition 3:} If hybrid power allocation algorithm is employed in HSRs, the average transmit power is independent with the velocity of train when transmission data rate requirements $(R_{di},R_{ds})$ are given. Namely, it is only determined by the deployment parameters of cellular network.}
\begin{proof}
See Appendix B.
\end{proof}

Based on these results, it can be concluded that the system parameter setting is independent with the velocity of train, which is very beneficial to the feasibility of applying this algorithm in HSRs.

\begin{table}
\caption{The minimum average transmit power under four different power allocation algorithms when $(R_{di},R_{ds})$ is given.}
\begin{tabular}{|c|c|c|c|c|}
\hline
Data Rate Pair $(R_{di},R_{ds})$ & FPA & WFA & CIA & HAA\\
(Mbps)   & (mW) & (mW) & (mW) & (mW) \\
\hline
( 20, 0 )  &   1713.6   &{\color{red}{928.3}}&   8343.7    &   {\color{red}{928.3}}  \\
\hline
( 15, 5 )  &   4732.1   &   21399.0  &   8343.7    &  {\color{red}{2064.1}}  \\
\hline
( 10,10 )  &  10360.0   &   27026.0  &   8343.7    &  {\color{red}{3654.3}}  \\
\hline
( 5, 15 )  &  17052.0   &   33719.0  &   8343.7    &  {\color{red}{5731.5}}  \\
\hline
( 0, 20 )  &  25010.4   &   41677.0  &{\color{red}{8343.7}}&  {\color{red}{8343.7}}  \\
\hline
\end{tabular}
\end{table}

\section{Performance Discussion in Non-uniform Motion Scenario}

For the case that high speed train travels along a line railway with a constant velocity, the transmission performance limit and corresponding optimal power allocation algorithm have been derived. However, it is very difficult to keep ideal uniform motion in practical system. The generalized case is that instantaneous traveling speed is always a stochastic process which varies along time within  a limited range. Thus, this section attempts to analyze the performance limits of hybrid information transmission in the non-uniform motion scenario, and discusses the performance loss of it compared with ideal uniform motion scenario. Firstly, we formulate the problem for non-uniform motion scenario in terms of achievable rate region, where some constraints that instantaneous velocity needs to meet are given. Though a universal case can not be obtained, we propose a worst case in terms of minimizing energy consumption to serve as the lower bound for system transmission performance in non-uniform motion scenario, and the performance loss is analyzed in terms of both achievable rate region and minimizing energy consumption.

\subsection{Formulating the Problem}

In HSR system, the whole train is accurately controlled by a servosystem. However, due to some non-ideal factors of system, the practical instantaneous speed of the train is usually time varying around the mean value so that $v(t)$ can be modeled as a stochastic process. Before formulating the problem, we need to consider the characteristics of $v(t)$, which will play an important role in the following discussion.

Provided that the mean value of $v(t)$ is $v_0$, most of all, $v(t)$ should be limited within a small range in normal running state, in order to guarantee the operation security. Secondly, the running time of each high speed train must be coincide with time table strictly. So it is reasonable to assume that the average time that the train takes to pass through the whole coverage of one BS is equal to $\frac{2L}{v_0}$, since driver has enough time to control the train to meet the operation time requirement during this period. Thus, considering the operation characteristics of practical HSRs, $v(t)$ under normal operation state must satisfy the following constraints.

\subsubsection*{Constraint 1}
The instantaneous speed $v(t)$ cannot exceed the maximum limit allowed by the system, which is expressed as
\begin{equation}\label{eqn:constraint 1}
  v(t)\in [v_0-\Delta v,v_0+ \Delta v]
\end{equation}
where $\Delta v$ denotes the maximum allowable shift of instantaneous speed, the value of which is usually very small compared with $v_0$. For example, the value of ratio $\Delta v /v_0$ in HSRs connecting Beijing and Shanghai in China is less than $1\%$.

\subsubsection*{Constraint 2}
The average speed during the coverage of each BS must be strictly equal to the mean value
\begin{equation}\label{eqn:constraint 2}
  \frac{v_0}{2L} \int_{-\frac{L}{v_0}}^{\frac{L}{v_0}} v(t) dt=v_0
\end{equation}

It is assumed that the system time is $t=-\frac{L}{v_0}$ in this case when the train passes through the point $E$ in Fig. \ref{fig:Base_Station_Deployment}. Since the average value of $v(t)$ is fixed to $v_0$, combining the above constraint in (\ref{eqn:constraint 2}), the system time period that the train travels from the point $E$ to point $D$ is still $[-\frac{L}{v_0},\frac{L}{v_0}]$. And the classification method for $(R_{di},R_{ds})$ presented in Section III is still reasonable in non-uniform motion scenario. Thus, we can still focus on the performance analysis during the period $t \in[-\frac{L}{v_0},\frac{L}{v_0}]$ from a perspective of achievable rate region, which can reflect all the characteristics of system. And the transmission distance at system time $t$ is
\begin{equation}\label{eqn:transmission distance in non-uniform case}
  d(t)=\sqrt{d_0^2+h_0^2+(\int_{-\frac{L}{v_0}}^t v(\tau)d \tau-L)^2},\,t\in \big[-\frac{L}{v_0},\frac{L}{v_0}\big]
\end{equation}

\begin{figure}[!t]
\centering
\includegraphics[width=3.5 in]{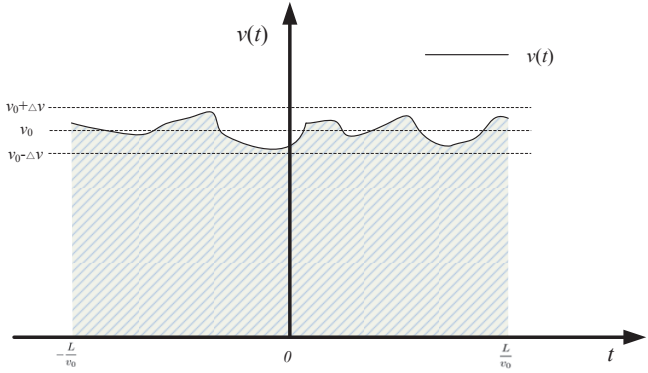}
\caption{An example for instantaneous speed realization $v(t)$ that meets the constraints in (\ref{eqn:constraint 1}-\ref{eqn:constraint 2}).}
\label{fig:velocity_distribution}
\end{figure}

Fig. \ref{fig:velocity_distribution} illustrates an example for the instantaneous speed realization $v(t)$, which is varying along time between the two extreme lines. The integration of $v(t)$ with respect to time during the period $t \in [-\frac{L}{v_0},\frac{L}{v_0}]$ in the shadow region is equal to $2L$ according to the constraint in (\ref{eqn:constraint 2}).

As a result, both $d(t)$ and $R(t)$ are stochastic functions with respect to $v(t)$. Let the set $\Gamma$ denote all possible distribution for the train's instantaneous speed $v(t)$ that meet the constraints in (\ref{eqn:constraint 1}-\ref{eqn:constraint 2}). In order to guarantee the QoS requirement, the problem in (\ref{equ:optimization problem for power allocation}) need to be rewritten as follows under this condition.
\begin{subequations}\label{equ:optimization problem for minimizing transmit power in non-uniform motion}
\begin{align}
   C_{R_{ds}}&=\max \limits_{p(t)} \Big\{\min\limits_{v(t)\in \Gamma} \big\{\frac{v_0}{2L}\int_{-\frac{L}{v_0}}^{\frac{L}{v_0}}[R(t)-R_{ds}]dt \big\}\Big\}   \tag{20}\\
   s.t.\,\,\,\,& R(t)=B \log_2\Big(1+\frac{G p(t)}{d(t)^\alpha \sigma_0^2}\Big) \label{equ:optimization problem for minimizing transmit power in non-uniform motion a} \\
   &  \frac{v_0}{2L}\int_{-\frac{L}{v_0}}^{\frac{L}{v_0}}p(t)dt  \leq P_0
     \label{equ:optimization problem for minimizing transmit power in non-uniform motion b} \\
   &  0 \leq R_{ds} \leq \min\limits_{v(t)\in \Gamma} \{R(t)\}, \, t\in \big[-\frac{L}{v_0},\frac{L}{v_0}\big]  \label{equ:optimization problem for minimizing transmit power in non-uniform motion c}
\end{align}
\end{subequations}

It is a typical random optimization problem. Though it is very difficult to obtain a universal results, we can investigate the lower bound performance of system by a deterministic worst case of $v(t)$, which will be introduced in the next subsection. Besides, it is worth noting that robust performance analysis in this section focuses on the mismatch between non-uniform motion and hybrid power allocation algorithm, not the performance loss caused by instantaneous speed estimation error. Since the transmission speed of feedback information from train is much faster than the variation rate of instantaneous speed, it is reasonable to assume that the control center at base station still can track the variation of instantaneous speed very well in this case.

\subsection{Performance Analysis in a Deterministic Worst Case}

As is shown in Fig. \ref{fig:velocity_distribution}, $v(t)$ can be arbitrary time-domain function that meets the above two constraints. This section will propose a deterministic worst case in terms of energy consumption, which can serve as a lower bound for the system performance of general case.

\emph{\textbf{Proposition 4:} The deterministic worst case for $v(t)$ realization that satisfies the constraints in (\ref{eqn:constraint 1}-\ref{eqn:constraint 2}) is a two-value function, which is expressed as the Eqn. (\ref{eqn:the speed in worst case}). Namely, HSR runs with the lowest allowable speed $v_0-\Delta v$ in the far region of base station coverage while runs with the highest allowable speed $v_0+\Delta v$ in the near region.}
\begin{equation}\label{eqn:the speed in worst case}
  v(t)=
\begin{cases}
   v_0-\Delta v, & \textrm{if $t \in [-\frac{L}{v_0},-\frac{L}{2v_0}] \cup [\frac{L}{2v_0},\frac{L}{v_0}]$},  \\
   v_0+\Delta v, & \textrm{if $t \in [-\frac{L}{2v_0},\frac{L}{2v_0}]$}.
\end{cases}
\end{equation}
\begin{proof}
See Appendix C.
\end{proof}

An intuitive proof of it is also given here to describe this problem more clear. Informally speaking, from a perspective of information transmission within a cell coverage, it is intuitive that the position on the railway that is far away from base station is more adverse for information transmission than the position that is near to base station on the railway due to great path loss. Since the total time that a train takes to pass through a base station coverage is fixed due to constraint (\ref{eqn:constraint 2}), it is better to spend more time in the region that is near to base station while spending less time in the region that is far from base station. Inversely, we can obtain a deterministic worst case of $v(t)$ realization for information transmission in HSRs, which is presented as Proposition 4.

Substituting the expression of the deterministic worst case shown in (\ref{eqn:the speed in worst case}) into the problem in (\ref{equ:optimization problem for minimizing transmit power in non-uniform motion}), the stochastic process $v(t)$ can be taken place of a deterministic one, which is very helpful for solving the problem in (\ref{equ:optimization problem for minimizing transmit power in non-uniform motion}). Similar to the results in Proposition 1, the conditional capacity function $C_{R_{ds}}$ in this case can be expressed as
\begin{equation}\label{eqn:conditional capacity function in non-uniform motion case}
\begin{split}
  C_{R_{ds}} & =\frac{v_0}{L}\int_{0}^{\frac{L}{2v_0}}\Big[B \log_2\big(1+\frac{G p_b^\ast(t)}{d_1(t)^\alpha \sigma_0^2}\big)\Big]dt  \\
  & +\frac{v_0}{L} \int_{\frac{L}{2v_0}}^{\frac{L}{v_0}}\Big[B \log_2\big   (1+\frac{G p_b^\ast(t)}{d_2(t)^\alpha \sigma_0^2}\big)\Big]dt-R_{ds}
\end{split}
\end{equation}
where
\begin{subequations}\label{eqn:the two distance expression}
\begin{align}
  & d_1(t)=\sqrt{d_0^2+h_0^2+[(v_0+\Delta v)t]^2} \\
  & d_2(t)=\sqrt{d_0^2+h_0^2+[(v_0-\Delta v)t+\tfrac{\Delta v}{v_0}L]^2} \\
  & p_b^\ast(t)=
\begin{cases}
  (\frac{d_1(t)^\alpha \sigma_0^2}{G}(2^{\frac{R_{ds}}{B}}-1),\frac{B}{\lambda_b}-\frac{d_1(t)^\alpha \sigma_0^2}{G})^+,t\in [0,\frac{L}{2v_0}]  \\
  (\frac{d_2(t)^\alpha \sigma_0^2}{G}(2^{\frac{R_{ds}}{B}}-1),\frac{B}{\lambda_b}-\frac{d_2(t)^\alpha \sigma_0^2}{G})^+,t\in [\frac{L}{2v_0},\frac{L}{v_0}]
\end{cases}
\end{align}
\end{subequations}
where $\lambda_b$ is determined by the constraint in (\ref{equ:optimization problem for minimizing transmit power in non-uniform motion b}).

\begin{figure}[!t]
\centering
\includegraphics[width=3.4 in]{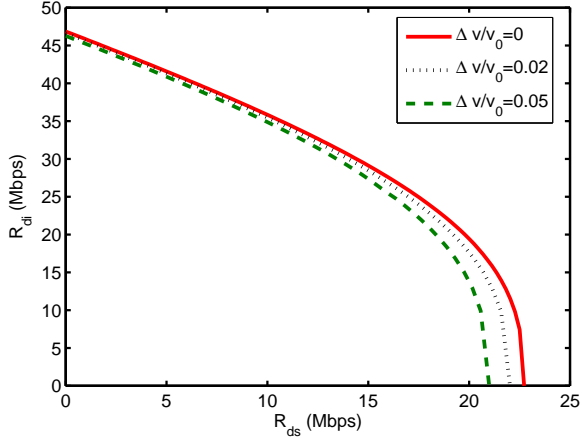}
\caption{The achievable rate regions in non-uniform motion when average transmit power $P_0$ is $40$ dBm, which contains three cases: $\Delta v/v_0$ is $0$, $0.02$ and $0.05$.}
\label{fig:Simulation_for_Two_Rate_Region_Nonideal}
\end{figure}

Based on above results, the QoS-distinguished achievable rate region $\{(R_{di},R_{ds})\}$ under non-uniform motion scenario can be derived. Fig. \ref{fig:Simulation_for_Two_Rate_Region_Nonideal} shows the corresponding achievable rate regions when average transmit power $P_0$ is $40$ dBm, which contains three cases: $\Delta v/v_0$ is $0$, $0.02$ and $0.05$. Obviously, the case that $\Delta v/v_0=0$ is equivalent to the uniform motion scenario that has been discussed in Section IV and Section V. It can be seen that non-uniform characteristic of instantaneous velocity can degrade the transmission performance in HSRs. However, the deterioration is not serious compared with ideal uniform motion scenario. Considering the fact that the value of $\Delta v/v_0$ is usually less than $0.01$ in practical system, such as the HSRs connection Beijing and Shanghai in China, it can be concluded that the effect of non-uniform motion on the transmission performance is very slight in terms of QoS-distinguished achievable rate region.

\begin{figure}[!t]
\centering
\includegraphics[width=3.4 in]{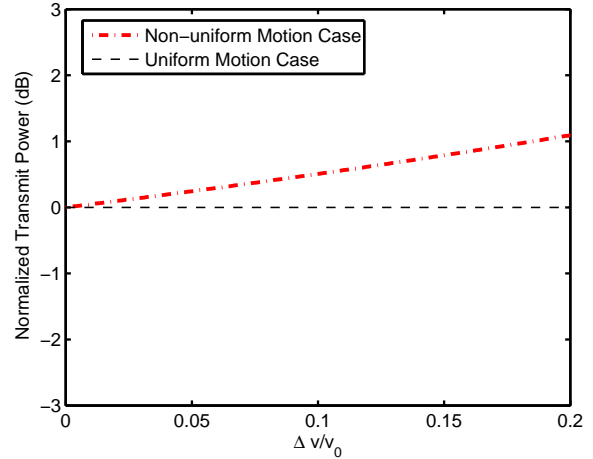}
\caption{The normalized minimum required average transmit power as a function of maximal velocity shift ratio with respect to that in uniform motion case, where data rate requirement pair $(R_{di},R_{ds})$ is $(30,10)$ Mbps.}
\label{fig:performance_loss_of_nonideal}
\end{figure}

In order to maintain the reliability of QoS requirements, the data rate requirements should be guaranteed even in worst case. It means that some power margin is essential to against the possible worst velocity realization case for non-uniform motion scenario, which can be regarded as the performance loss caused by non-ideal uniform motion of train. From a perspective of minimizing transmit power on the condition that data rate requirement $(R_{di},R_{ds})$ is given, Fig. \ref{fig:performance_loss_of_nonideal} shows the normalized minimum required transmit power as a function of maximal velocity shift ratio $\Delta v / v_0$ with respect to that in uniform motion scenario, where $(R_{di},R_{ds})$ is equal to $(30,10)$ Mbps. The baseline of uniform motion case is also shown to serve as a comparison. As observed from Fig. \ref{fig:performance_loss_of_nonideal}, the performance loss caused by non-ideal uniform motion can be limited within $1$dB even when the velocity shift ratio is approach to $0.2$. As we know, the value of $\Delta v/v_0$ under normal running status in practical system is strictly controlled in the range of $0.01$ due to security consideration. Thus, performance loss is further small within the acceptable range.

\section{Conclusion}

In HSRs wireless communication scenario, the information flows transmitted between base station and high speed train are hybrid version which have diverse QoS requirements. Neither throughput nor outage capacity is sufficient to describe this kind of information transmission scenario. As a result, this paper concentrated on the performance limits of hybrid information transmission in HSRs scenario.
Based on delay requirements, these information flows are divided into two categories: delay-insensitive information and delay-sensitive information. \emph{QoS-distinguished achievable rate region} was utilized to characterize transmission performance of these two concurrent streams. By solving the optimization problem, the expression of achievable rate region has been obtained with the help of conditional capacity function. And the corresponding optimal power allocation strategy was also derived, which can achieve the largest boundary of achievable rate region compared with other conventional algorithms. Then, from a view of green communication, the QoS-distinguished power allocation algorithm was given in terms of minimizing energy consumption when data rate pair is given. Lastly, we discussed the performance of hybrid information transmission in non-uniform motion scenario in terms of achievable rate region. The performance loss compared with uniform motion scenario was also indicated by simulation results.
Besides, it is worth noting that the QoS-distinguished achievable rate region in this paper is not limited to the HSRs scenario. It can be extended to the system performance over arbitrary time-varying channel scenarios.

\appendices
\section{Proof of Proposition 1}

From the systematic level, the transmit power can be divided into two parts: one for delay-sensitive information stream and the other for delay-insensitive information stream, which are denoted as $p_1(t)$ and $p_2(t)$, respectively.
\begin{equation}\label{eqn:relationship of transmit power}
  p(t)=p_1(t)+p_2(t),t\in [-\frac{L}{v_0},\frac{L}{v_0}]
\end{equation}

Since $R_{ds}$ is the delay-sensitive information rate that must be satisfied with priority, the expression of $p_1(t)$ can be derived in order to avoid outage event.
\begin{equation}\label{eqn:transmit power for delay-sensitive information}
  p_1(t)=\frac{d(t)^\alpha \sigma_0^2}{G}(2^{\frac{R_{ds}}{B}}-1),t\in [-\frac{L}{v_0},\frac{L}{v_0}]
\end{equation}

Then, according to the Lagrange principle, we can establish the following function
\begin{equation}\label{eqn:Largrange function}
\begin{split}
  F=& \frac{v_0}{2L}\int_{-\frac{L}{v_0}}^{\frac{L}{v_0}}B \log_2 \Big(1+\frac{G (p_1(t)+p_2(t))}{d(t)^\alpha \sigma_0^2}\Big)dt \\
    & -R_{ds} -\lambda_0' \cdot \frac{1}{2L/v_0}\int_{-\frac{L}{v_0}}^{\frac{L}{v_0}}(p_1(t)+p_2(t))dt
\end{split}
\end{equation}

By setting the first-order derivation of function $F$ with respect to $p_2(t)$ to equal zero, we can get the expression of $p_2(t)$, which is expressed as
\begin{equation}\label{eqn:expression of p2}
  p_2(t)=\Big(\frac{B}{\lambda_0'}-\frac{d(t)^\alpha \sigma_0^2}{G}-p_1(t),0\Big)^+,t\in [-\frac{L}{v_0},\frac{L}{v_0}]
\end{equation}
where operation $(x_1,x_2)^+$ denotes the larger one between $x_1$ and $x_2$. $\lambda_0'$ is a constant value which is determined by the average power constraint in (\ref{equ:optimization problem for power allocation a})

Combining the results in (\ref{eqn:transmit power for delay-sensitive information}) and (\ref{eqn:expression of p2}), the corresponding optimal power allocation strategy is
\begin{equation}\label{equ:optimal power allocation solution in appendix}
  p_{hb}^\ast(t)=\Big(\frac{d(t)^\alpha \sigma_0^2}{G}(2^{\frac{R_{ds}}{B}}-1),\frac{B}{\lambda_0'}-\frac{d(t)^\alpha \sigma_0^2}{G}\Big)^+,t\in [-\frac{L}{v_0},\frac{L}{v_0}]
\end{equation}

Substituting (\ref{equ:optimal power allocation solution in appendix}) into (\ref{equ:optimization problem for power allocation}), the expression of conditional capacity function can be obtained
\begin{equation}\label{equ:maximum non-real-time rate in appendix}
  C_{R_{ds}}=\frac{v_0}{2L}\int_{-\frac{L}{v_0}}^{\frac{L}{v_0}}\Big[B \log_2\big(1+\frac{G p_{hb}^\ast(t)}{d(t)^\alpha \sigma_0^2}\big)-R_{ds}\Big]dt
\end{equation}

Thus, the Proposition 1 has been proved.

\section{Proof of Proposition 3}

It is obvious that the average transmit power can be calculated by the Eqn. (\ref{equ:optimal power allocation solution in terms of minimizing transmit power})-(\ref{eqn:minimum transmit power in Proposition 2}) when the data rate pair $(R_{di},R_{ds})$ is given. Then, through a variable substitution $v_0 t=L x$, and using $\frac{Lx}{v_0}$ to take place of $t$, we can rewritten the Equations (\ref{equ:optimal power allocation solution in terms of minimizing transmit power})-(\ref{eqn:minimum transmit power in Proposition 2}) as:
\begin{equation}\label{eqn:rewritten expressions in Proposition 2-1}
  p_{m}^\ast(x)=\Big(\frac{d(x)^\alpha \sigma_0^2}{G}(2^{\frac{R_{ds}}{B}}-1), \frac{B}{\lambda_0''}-\frac{d(x)^\alpha \sigma_0^2}{G}\Big)^+
\end{equation}
\begin{equation}\label{eqn:rewritten expressions in Proposition 2-2}
  R_{di}=\int_{0}^{1}\Big[B \log_2\big(1+\frac{G p_{m}^\ast(x)}{d(x)^\alpha \sigma_0^2}\big)-R_{ds}\Big]dx
\end{equation}
\begin{equation}\label{eqn:rewritten expressions in Proposition 2-3}
  P_{0}^{min}=\int_{0}^{1}p_m^\ast(x)dx\,\,\,\,\,
\end{equation}
where
\begin{equation}\label{eqn:rewritten the expression of distance}
  d(x)^\alpha=[d_0^2+h_0^2+(Lx)^2]^{\frac{\alpha}{2}}
\end{equation}

It can be seen that the variable $v_0$ is not contained in the equation set shown in (\ref{eqn:rewritten expressions in Proposition 2-1})- (\ref{eqn:rewritten the expression of distance}). As a consequence, the average transmit power $P_{0}^{min}$ is independent with the velocity of train, and is only determined by deployment parameters.

Thus, the Proposition 3 has been proved.

\section{Proof of Proposition 4}

Firstly, it can be confirmed that the expression of instantaneous speed in (\ref{eqn:the speed in worst case}) satisfies the constraints in (\ref{eqn:constraint 1}) and (\ref{eqn:constraint 2}). Then, we will prove the worst property of it by contradiction in terms of maximizing energy consumption.

It is a typical two-value piecewise function. The whole period can be divided into two phases: the phase one is with velocity $v_0+\Delta v$ and the phase two is with velocity $v_0-\Delta v$, the demarcation point of which is $\frac{v_0+\Delta v}{v_0} \frac{L}{2}$ away from the point $O$ in Fig. \ref{fig:Base_Station_Deployment}. The total energy consumption during the period $[-\frac{L}{v_0},\frac{L}{v_0}]$ can be expressed as
\begin{equation}\label{eqn:total energy consumption}
  E=2\int_{0}^{\frac{L}{2v_0}} p(l(v+\Delta v,t))dt+ 2\int_{\frac{L}{2v_0}}^{\frac{L}{v_0}} p(l(v-\Delta v,t))dt
\end{equation}
where $l(v+\Delta v,t)$ denotes the transmission distance between BS and AP when velocity is $v+\Delta v$ at system time $t$, and $p(l(v+\Delta v,t))$ denotes the instantaneous transmit power.

Assuming the velocity $v_0+\Delta v$ is not optimal for the first phase. Due to the constraint in (\ref{eqn:constraint 1}), the optimal value $v(t)$ may be arbitrary time varying one or time non-varying one that is less than $v_0+\Delta v$. As a result, due to the decrement of velocity, the time range of the first phase will become large, which is denoted as $\frac{L}{v_0}+\Delta t$. Since the total time that the train travels from $E$ to $D$ is limited to $\frac{2L}{v_0}$ in (\ref{eqn:constraint 2}), the time that the second phase experience should be $\frac{L}{v_0}-\Delta t$. Now we calculate the total energy consumption under above assumption.
\begin{equation}\label{eqn:total energy consumption in assumption}
  E'=2\int_{0}^{\frac{L}{2v_0}+\Delta t} p(l(v(t),t))dt+ 2\int_{\frac{L}{2v_0}+\Delta t}^{\frac{L}{v_0}}p(l(v-\Delta v,t))dt
\end{equation}

Comparing (\ref{eqn:total energy consumption}) with (\ref{eqn:total energy consumption in assumption}), it can be observed that the train will spend more time in the position that is close to the base station and spend less time in the position that is far from base station. Thus, energy consumption will become less, namely $E' \leq E$. So the velocity should be equal to $v_0+\Delta v$ in phase one in terms of maximizing energy consumption, to establish a worst case.

Similar results can be derived for the phase two. Thus, the expression in (\ref{eqn:the speed in worst case}) is the worst case. The Proposition 4 has been proved.

\section*{Acknowledgment}

The authors would like to thank professor Nosratinia Aria for his insightful comments and suggestions. This work was partly supported by the China Major State Basic Research Development Program (973 Program) No.2012CB316100(2).

\ifCLASSOPTIONcaptionsoff
  \newpage
\fi

\end{document}